\documentstyle[epsfig,natbib2,natbibmnfix]{mn}

\newcommand{\be}{\begin{equation}}
\newcommand{\ee}{\end{equation}}

\def\ltsima{$\; \buildrel < \over \sim \;$}
\def\simlt{\lower.5ex\hbox{\ltsima}}
\def\gtsima{$\; \buildrel > \over \sim \;$}
\def\simgt{\lower.5ex\hbox{\gtsima}}

\title{The Diversity and Similarity of Simulated Cold Dark Matter Halos}

\author[Navarro et al..] {\parbox{18cm}{
Julio F. Navarro$^{1,5}$,
Aaron Ludlow$^{1},$ 
Volker~Springel$^{2}$, 
Jie~Wang$^{2}$,\\ 
Mark Vogelsberger$^{2}$,
Simon D.M. White$^{2}$, 
Adrian Jenkins$^{3}$, 
Carlos S. Frenk$^{3}$, 
and Amina Helmi$^{4}$,
}\vspace{0.3cm}\\
$^{1}${Dept. of Physics and Astronomy, University of
    Victoria, Victoria, BC, V8P 5C2, Canada}\\
$^2$Max-Planck-Institut f\"{u}r Astrophysik,
Karl-Schwarzschild-Stra\ss{}e 1, 85740 Garching bei M\"{u}nchen,
Germany\\
$^{3}${Institute for Computational Cosmology, Dep. of Physics, Univ. of
  Durham, South Road, Durham  DH1 3LE, UK}\\
$^{4}${Kapteyn Astronomical Institute, Univ. of Groningen,
P.O. Box 800, 9700 AV Groningen, The Netherlands}\\
$^{5}${Department of Astronomy, University of Massachusetts, Amherst,
    MA 01003-9305, USA}\\
}

\begin{document}

\maketitle 
\begin{abstract}
We study the mass, velocity dispersion, and anisotropy profiles of
$\Lambda$CDM halos using a suite of N-body simulations of
unprecedented numerical resolution. The {\it Aquarius Project} follows
the formation of 6 different galaxy-sized halos simulated several
times at varying numerical resolution, allowing numerical convergence
to be assessed directly.  The highest resolution simulation represents
a single dark matter halo using 4.4 {\it billion} particles, of which
1.1 billion end up within the virial radius. Our analysis confirms a
number of results claimed by earlier work, and clarifies a few issues
where conflicting claims may be found in the recent literature. The
mass profile of $\Lambda$CDM halos deviates slightly but
systematically from the form proposed by Navarro, Frenk \& White. The
spherically-averaged density profile becomes progressively shallower
inwards and, at the innermost resolved radius, the logarithmic slope
is $\gamma \equiv -$d$\ln\rho/$d$\ln r \simlt 1$.  Asymptotic inner
slopes as steep as the recently claimed $\rho \propto r^{-1.2}$ are
clearly ruled out.  The radial dependence of $\gamma$ is well
approximated by a power-law, $\gamma \propto r^{\alpha}$ (the Einasto
profile). The shape parameter, $\alpha$, varies slightly but
significantly from halo to halo, implying that the mass profiles of
$\Lambda$CDM halos are not strictly universal: different halos cannot,
in general, be rescaled to look identical. Departures from
similarity are also seen in velocity dispersion profiles and correlate
with those in density profiles so as to preserve a power-law form for
the spherically averaged pseudo-phase-space density,
$\rho/\sigma^3\propto r^{-1.875}$. The index here is identical to that
of Bertschinger's similarity solution for self-similar infall onto a
point mass from an otherwise uniform Einstein-de Sitter Universe. The
origin of this striking behaviour is unclear, but its robustness
suggests that it reflects a fundamental structural property of
$\Lambda$CDM halos. Our conclusions are reliable down to radii below
$0.4\%$ of the virial radius, providing well-defined predictions for
halo structure when baryonic effects are neglected, and thus an
instructive theoretical template against which the modifications
induced by the baryonic components of real galaxies can be judged.
\end{abstract}


\section{Introduction}
\label{intro}

\renewcommand{\thefootnote}{\fnsymbol{footnote}}
\footnotetext[1]{E-mail: jfn@uvic.ca}

A couple of decades of steady progress in the simulation of non-linear
structures in a cold dark matter (CDM) dominated universe have
resulted in significant advances in our understanding of the
clustering of dark matter on the scale of galactic halos. There is now
widespread consensus that the hierarchical assembly of CDM halos
yields: (1) mass profiles that are approximately ``universal'' (i.e.,
independent of mass and cosmological parameters aside from simple
physical scalings \citep[][hereafter NFW]{Navarro1996,Navarro1997},
(2) strongly triaxial shapes, with a slight preference for nearly
prolate systems \citep[e.g.,][]{Frenk1988,JingSuto2002,Allgood2006,Hayashi2007},
(3) abundant, but non-dominant, substructure
\citep[]{Klypin1999,Moore1999b,Ghigna2000,Gao2004}, and (4) ``cuspy''
inner mass profiles, where the central density increases
systematically as the numerical resolution of the calculation is
improved \citep[see, e.g., NFW,][]{Moore1999a,Fukushige2001,Navarro2004,Diemand2005}.

Despite this consensus, there are a number of issues where conflicting
claims may be found in the recent literature, hindering the design and
interpretation of observational tests aimed at validating or ruling
out various aspects of the CDM theory on these scales. One contentious
issue concerns the statistics, spatial distribution, and
structure of substructure, and their consequences for the discovery
and interpretation of possible signals of dark matter annihilation in
the gamma-ray sky \citep[][and references
therein]{Stoehr2003,Diemand2007,Kuhlen2008,AqNatPaper}.  The
controversy extends to the structure of the inner cusps both of the main
halo and of substructure halos, where some recent work has claimed a
well-defined central slope of $\rho\propto
r^{-1.2}$\citep[]{Diemand2004,Diemand2005,Diemand2008} whereas others
have argued that no compelling evidence for such power-law behaviour
is apparent \citep{Navarro2004,Graham2006}.

Considerable debate also surrounds whether the structure of CDM halos
is truly ``universal''.  This is indeed the case if halos have mass
profiles that are well-described by two-parameter formulae, such as
the NFW profile or some of its modifications; see, for example,
\citet[][hereafter M99]{Moore1999a}. These profiles have two scaling
parameters (mass and size) but fixed {\it shape}, so that two
different halos can, in principle, be rescaled to be indistinguishable
from each other.

On the other hand, recent work suggests that at least {\it three}
parameters may be needed to describe halo mass profiles accurately. An
example is the Einasto formula \citep[]{Einasto1965}, shown by
\citet{Navarro2004} to improve significantly the accuracy of the fits
to the inner density profiles of simulated halos. It is unclear from
that work, however, whether the improvement is due to the fact that
the Einasto formula has a different asymptotic inner behaviour than
NFW or to the extra shape parameter it
introduces. \citet{Merritt2005,Merritt2006} explored this further and
argued that the third parameter is indeed needed to account faithfully
for the curvature in the shape of the density profile. Merritt et
al.'s conclusions have received support from the work of
\citet{Gao2008} and \citet{Hayashi2008}, who have stacked density
profiles of many halos of similar mass to show that mean profile
shape, and, in particular, the Einasto shape parameter $\alpha$ (see
eq.~\ref{eq:einasto} below), depends systematically on halo mass. This
implies that the mass profile of $\Lambda$CDM halos is not strictly
universal; no simple scaling of the average profile of cluster halos
will provide an accurate fit to the average profile of galaxy halos.

Many of these controversies and uncertainties may be traced to the
fact that earlier work has lacked the numerical resolution and the
representative halo sample needed to
settle the debate. For example, the dark matter annihilation flux
observable from Earth depends crucially on resolving not only
substructures but also the nested ``substructure within substructure''
expected from the hierarchical assembly of CDM halos. Only the most
recent simulations have been able to begin addressing this issue
\citep[see, e.g.,][]{Diemand2008,AqNatPaper,AqPaper2}.  

A similar comment applies to the structure of the inner cusp, where
pinning down the asymptotic inner behaviour of the dark matter density
profile depends crucially on understanding the limitations introduced
by, for example, finite particle number, gravitational
softening, and time-stepping technique.

We have shown in earlier work \citep[][hereafter P03]{Power2003} that,
when suitable choices of the numerical parameters are made, the main
factor determining the innermost radius where the mass profile may be
measured reliably is the total number of particles used in the
simulation.  Empirically, the boundary of the region where numerical
convergence is achieved roughly corresponds to the radius where the
two-body relaxation time, $t_{\rm relax}$, exceeds the age of the
Universe. Since $t_{\rm relax}$ scales roughly like the enclosed
number of particles times the local orbital timescale, and the latter
drops sharply toward the centre, extending the resolved region inwards
even modestly requires a dramatic increase in the total number of
particles.

These difficulties, coupled to the significant halo-to-halo scatter
already seen in early work, imply that substantive progress on these
issues requires a concerted numerical effort where {\it several}
different halos are simulated with {\it varying numerical resolution},
so that cosmic variance and numerical convergence may be assessed
directly.

These are the aims of {\it The Aquarius Project}, a recently completed
suite of numerical simulations of the formation of galaxy-sized halos
in the $\Lambda$CDM cosmogony. The series includes re-simulations of
six different $\sim 10^{12} \, M_{\odot}$ halos where the number of
particles is systematically varied. In one case, the {\it same} halo
is simulated 5 times, increasing gradually the number of particles in
the halo from about one million to $\sim 1.1$ {\it billion} within the
virial radius. The highest resolution simulations of the other 5 halos
have roughly $100$-$200$ million particles each within the virialized
region.

The simulation series has been presented recently by
\citet[]{AqNatPaper,AqPaper2}, where the interested reader may
find relevant details. Our first paper \citep[]{AqNatPaper} deals
with predictions of the annihilation signal whereas the second
\citep[]{AqPaper2} addresses the statistics, spatial
distribution, and structure of dark matter substructures. Here we deal
with the structure of the {\it main halo}, with special emphasis on
the structure of the inner cusp. The plan of the paper is as
follows. Sec.~\ref{sec:numexp} summarizes briefly the numerical
parameters of our simulations; Sec.~\ref{sec:mprof} and
~\ref{sec:dynprof} present our main results. We conclude with a brief
discussion and summary in Sec.~\ref{sec:conc}.


\begin{figure*}
\begin{center}
\resizebox{17cm}{!}{\includegraphics{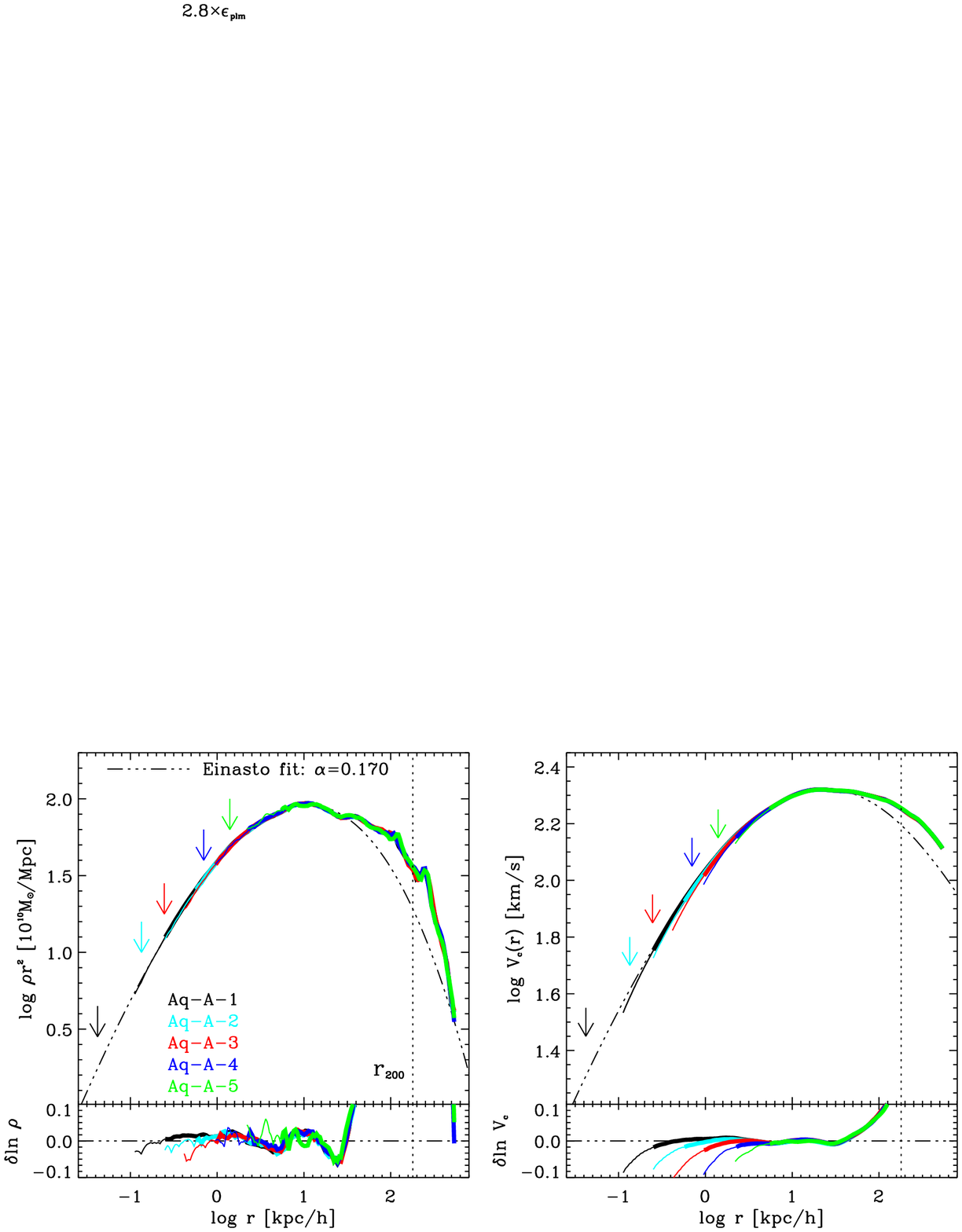}}
\end{center}
\caption{Spherically-averaged density (left) and circular velocity (right) profiles for the
  Aq-A halo simulation series. Different colours correspond to
  different resolution runs, as labeled in the figure. The density
  profile is multiplied by $r^2$ in order to emphasize small
  deviations. The bumps in the outer regions may be traced to the
  presence of substructure and unrelaxed tidal debris. Profiles are
  shown from $\sim 3 r_{200}$ down to the ``convergence radius'',
  $r_{\rm conv}^{(1)}$, corresponding to the radius where the
  relaxation time, $t_{\rm relax}$, is of the order of the age of the
  Universe.  The thick portion of each profile indicates the region
  $r>r_{\rm conv}^{(7)}$ where $t_{\rm relax}$ is more than 7 times
  the age of the universe and where stricter convergence is
  achieved. Outside $r_{\rm conv}^{(7)}$ circular velocity estimates
  converge to better than $2.5\%$ (see Fig.~\ref{FigConvVc}). The
  dot-dashed line shows an Einasto profile with $\alpha=0.17$ matched
  at ($r_{-2}$,$\rho_{-2}$), the peak in the $r^2\rho$ profile. This
  provides an excellent fit to the structure of the inner regions of
  the halo, as shown by the residuals plotted in the bottom
  panels. Arrows indicate the softening length $h_s$ of each
  simulation.\label{FigRhoVc}}
\end{figure*}

\begin{figure}
\begin{center}
\resizebox{8cm}{!}{\includegraphics{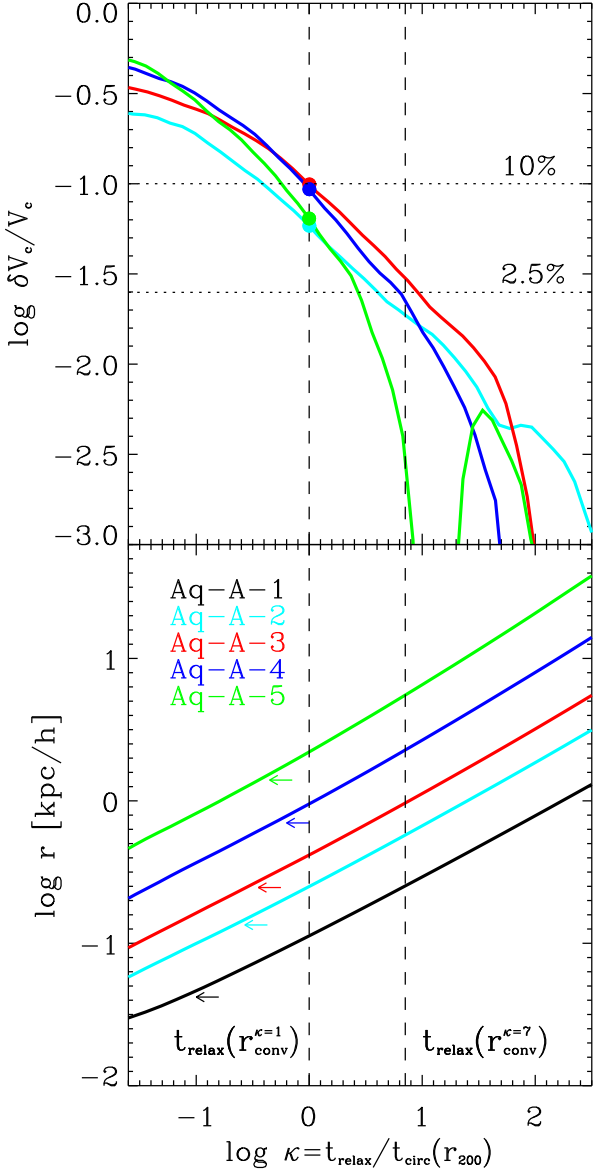}}
\end{center}
\caption{
{\it Top panel:} Fractional deviations in the circular velocity
profile of the Aq-A convergence series versus the (enclosed)
relaxation time, $t_{\rm relax}$, expressed in units of the circular
orbit period at the virial radius, $t_{\rm
circ}(r_{200})$. Deviations are measured relative to the highest
resolution halo, Aq-A-1. Note that departures from convergence for all
simulations are similar when expressed this way, indicating that
$t_{\rm relax}$ is the main parameter determining convergence. Solid
circles mark the location of the convergence criterion proposed by
P03. Note that $V_c$ estimates converge there to about
10\%. A stricter convergence criterion, e.g., $2.5\%$ convergence in
$V_c$, is achieved at larger radii, where $t_{\rm relax}\sim 7 \,
t_{\rm circ}(r_{200})$ (right vertical line). {\it Bottom panel:}
Relaxation time versus radius for all five Aq-A simulations. Arrows
indicate $h_s=2.8\, \epsilon_G$, the lengthscale where 
pairwise interactions become Newtonian.
\label{FigConvVc}}
\end{figure}

\section{The Numerical Simulations} \label{sec:numexp}

We present here for completeness a brief summary of the numerical
simulations, and refer the reader to
\citet[]{AqNatPaper,AqPaper2} for further details.

\subsection{The Cosmological Parameters}

All our simulations assume a $\Lambda$CDM cosmogony with the following
parameters: $\Omega_{\rm m} = 0.25$, $\Omega_\Lambda=0.75$, $\sigma_8=0.9$,
$n_s=1$, and Hubble constant $H_0 =100 \,h\,{\rm km\,s^{-1}\,Mpc^{-1}}
= 73\,{\rm km\,s^{-1}\,Mpc^{-1}}$. These cosmological parameters are
the same adopted in previous numerical work by our group, such as the
Millennium Simulation of \citet{Springel2005a}, and are consistent,
within their uncertainties, with constraints derived from the WMAP 1-
and 5-year data analyses \citep[]{Spergel2003,Komatsu2008} and
with the recent cluster abundance analysis of \citet[]{Henry2008}.

\subsection{The Code}

The simulations were carried out with a new version of the {\small
GADGET} \citep[]{Springel2001a, Springel2005b} parallel cosmological
code. This version, which we call {\small GADGET-3}, has been
especially developed for this project, and implements a novel domain
decomposition technique in order to achieve unprecedented dynamic
range in massively-parallel computer systems without sacrificing load
balancing or numerical accuracy.  Time stepping is carried out with a
kick-drift-kick leap-frog integrator where the timesteps are based on
the local gravitational acceleration, together with a conservatively
chosen maximum timestep for all particles.

Pairwise particle interactions are softened with a spline of
scalelength $h_s$, so that they are strictly Newtonian for particles
separated by more than $h_s$. The resulting softening is roughly
equivalent to a traditional Plummer-softening with scalelength
$\epsilon_G\sim h_s/2.8$. The gravitational softening length is kept
fixed in comoving coordinates throughout the evolution of all our
halos. The dynamics is then governed by a Hamiltonian and the
phase-space density of the discretized particle system should be
strictly conserved as a function of time \citep{Springel2005b}.

\subsection{Halo Selection}

All halos in the Aquarius suite were identified for resimulation in a
$900^3$-particle parent simulation of a $100 \, h^{-1}$Mpc box. The
identification technique selects all $\sim 10^{12} \, M_{\odot}$ halos
in the box and chooses, at random, a few of them that satisfy a mild
isolation criterion (no neighbour exceeding half its mass within $1
h^{-1}$ Mpc). This criterion is only imposed in order to remove halos
in the vicinity of massive groups and clusters, which may have evolved
differently from the average.

Each halo is then resimulated at various resolutions, making sure that
each resimulation shares the same power spectrum and phases at all
resolved spatial frequencies. Initial displacements are imprinted
using the Zeldovich approximation and a `glass-like' uniform particle
load \citep{White1996}. The $100 \, h^{-1}$Mpc simulation box is
divided into a ``high-resolution'' region, which corresponds to the
Lagrangian region surrounding the target halo, and a low-resolution
region (the rest of the box), which is represented with a smaller
number of particles with mass increasing with distance to the target
halo. We have carefully designed the geometry of the high-resolution
region in order to avoid contamination of the halo by massive
low-resolution particles. Typically, about 30\% of particles in the
high-resolution region end up in the virialized region of the final
halo, and {\it no} higher mass particles end up within the virial
radius of the final halo.

\begin{figure*}
\begin{center}
\resizebox{18cm}{!}{\includegraphics{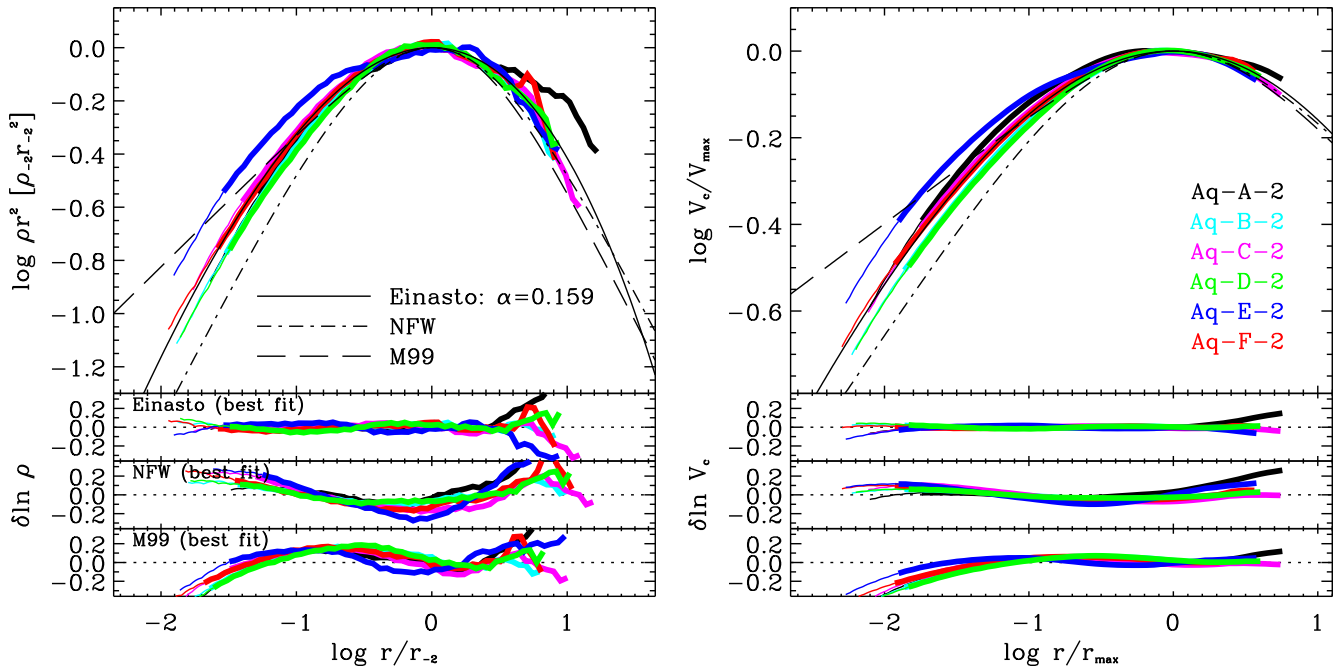}}
\end{center}
\caption{ {\it Left:} Spherically-averaged density profiles of all
  level-2 Aquarius halos. Density estimates have been multiplied by
  $r^2$ in order to emphasize details in the comparison. Radii have
  been scaled to $r_{-2}$, the radius where the logarithmic slope has
  the ``isothermal'' value, $-2$. Thick lines show the profiles from
  $r_{\rm conv}^{(7)}$ outward; thin lines extend inward to $r_{\rm
  conv}^{(1)}$. For comparison, we also show the NFW and M99 profiles,
  which are fixed in these scaled units. This scaling makes clear that
  the inner profiles curve inward more gradually than NFW, and are
  substantially shallower than predicted by M99. The bottom panels
  show residuals from the {\it best fits} (i.e., with the radial
  scaling free) to the profiles using various fitting formulae
  (Sec.~\ref{ssec:fitform}). Note that the Einasto formula fits all
  profiles well, especially in the inner regions. The shape parameter,
  $\alpha$, varies significantly from halo to halo, indicating that
  the profiles are not strictly self-similar: no simple physical
  rescaling can match one halo onto another. The NFW formula is also
  able to reproduce the inner profiles quite well, although the slight
  mismatch in profile shapes leads to deviations that
  increase inward and are maximal at the innermost resolved point. The
  steeply-cusped Moore profile gives the poorest fits. {\it Right:}
  Same as left, but for the circular velocity profiles, scaled to
  match the peak of each profile. This cumulative measure removes the
  bumps and wiggles induced by substructures and confirms the lack of
  self-similarity apparent in the left panel.
\label{FigDensProf}}
\end{figure*}

Table~\ref{tab:AqSims} lists some basic information about each
simulation. This includes a symbolic simulation name, the particle
mass in the high-resolution region, $m_p$, the gravitational softening
length, $\epsilon_G$, the virial radius\footnote{We define the virial
mass of a halo, $M_{200}$, as that contained within a sphere of mean
density $200\times \rho_{\rm crit}$. The virial mass defines
implicitly the virial radius, $r_{200}$, and virial velocity,
$V_{200}=(GM_{200}/r_{200})^{1/2}$, of a halo, respectively. We note
that other definitions of ``virial radius'' have been used in the
literature; the most popular of the alternatives adopts a density
contrast (relative to critical) of $\Delta\approx 178 \, \Omega_{\rm
m}^{0.45}\sim 100$ (for our adopted cosmological parameters, see
\citet{Eke1996}). We shall refer to these alternative choices, where
appropriate, with a subscript indicating the value of $\Delta$; i.e.,
$r_{50}$ would be the virial radius obtained assuming $\Delta=50$, and
so an enclosed density 200 times the {\it mean} cosmic value.},
$r_{200}$, as well as the total mass, $M_{200}$ and the total number
of particles, $N_{200}$, enclosed within $r_{200}$. Other structural
parameters of interest include the location of the peak in the
circular velocity profile, specified by $r_{\rm max}$ and $V_{\rm
max}$, as well as that of the velocity dispersion profile
($\sigma_{\rm max}$ and $r(\sigma_{\rm max})$). $\sigma_{\rm host}$
indicates the 1D rms velocity of the main halo within $r_{200}$
(excluding substructures).

Table~\ref{tab:AqSims} lists only information on the halos used in
this paper. A more complete list of numerical parameters may be found
in \citet[]{AqPaper2}. One of our halos, labeled Aq-A, has been
resimulated 5 times, spanning a factor of $\sim 2000$ in particle
mass. Our naming convention uses the tags ``Aq-A'' through ``Aq-F'' to
refer to each of the six Aquarius halos. An additional suffix ``1'' to
``5'' denotes the resolution level. ``Aq-A-1'' is our highest
resolution calculation: it follows the surroundings of Aq-A with $\sim
4.4$ billion particles, $\sim 1.1$ billion of which end up within
$r_{200}$. We have level-2 simulations of all 6 halos, corresponding
to between $100$ and $200$ million particles per halo (within
$r_{200}$).  The softening parameters of each simulation adopt the
``optimal'' softening recommendation of P03, which aims to balance the
number of timesteps required for accurate integration whilst minimizing
the loss of spatial resolution.

\subsection{Radial Profiles}
\label{ssec:prof}

Our analysis uses spherically-averaged profiles of the basic dynamical
properties describing the structure of $\Lambda$CDM halos: the
density, circular velocity, velocity dispersion, and anisotropy
profiles. Typically, these are computed in $50$ spherical shells
equally spaced in $\log_{10}r$ (where $r$ is the distance to the halo
center), and spanning the range $1.5\times 10^{-4}
<r/r_{200}<3$. (When different choices for either the number of bins
or the radial range are made, this is stated explicitly in the
analysis below.) These concentric shells are centered at the location
of the particle identified by the {\small SUBFIND} algorithm
\citep{Springel2001a} as having the minimum gravitational
potential. Extensive tests show that this procedure identifies the
region where the local density of the main subsystem of each halo
peaks, and is coincident in most cases (except perhaps major ongoing
mergers between comparable-mass halos) with the results of other
methods, such as the ``shrinking sphere'' method discussed by P03.

The mass density in each radial bin is estimated as the dark mass in
the bin divided by its volume, and assigned to a radius corresponding
to the bin center. Circular velocities are computed by adding up the
mass of each bin plus all interior ones, and assigned to the radius
corresponding to the outer edge of the bin. The construction of
velocity dispersion and anisotropy profiles is described in detail in
Sec.~\ref{ssec:velprof}. When differentiation is necessary, such as
when computing the logarithmic slopes shown in
Figs.~\ref{FigDlnrhoDlnr} and \ref{FigDlnrhoDlnrProf}, we use a simple
3-point Lagrangian interpolation to perform the numerical
differentiation (as implemented by the {\small DERIV} subroutine of
the {\small IDL} software package).

\section{Mass Profiles}
\label{sec:mprof}

\subsection{Numerical Convergence}

We begin our study of the mass profile by using our series of
re-simulations of the Aq-A halo in order to assess the radial range
where numerical convergence is achieved. Figure~\ref{FigRhoVc} shows
the mass profile of the five Aq-A resimulations; the left panels show
the spherically-averaged density profile (multiplied by $r^2$ in order
to emphasize small departures); the right panels the corresponding
circular velocity profile. Lines of different colours correspond to
different resimulations, as labeled. Arrows indicate $h_s=2.8\,
\epsilon_G$, the lengthscale where softened pairwise interactions
become fully Newtonian.

This figure demonstrates the striking numerical convergence achieved
in our re-simulations. Outside some characteristic radius (which we
discuss below), all the profiles are essentially indistinguishable
from each other, even down to details such as ``bumps'' in the outer
regions caused by the presence of substructure. As discussed by
\citet[]{AqPaper2}, this reflects the high quality of the
numerical integration of {\small GADGET-3} and the careful approach we
have taken to building our initial conditions; indeed, the Aq-A
resimulations not only reproduce faithfully the properties of the main
halo, but even the mass, location and internal structure of most major
substructures.
 
Inevitably, near the centre the mass profiles diverge as a consequence
of numerical limitations.  Each profile is plotted down to the
``convergence radius'' proposed by P03. These authors demonstrate that deviations from convergence depend (for appropriate choices
of other numerical parameters) solely on the number of particles, and
scale roughly with the collisional ``relaxation'' time, $t_{\rm
relax}$. Expressed in units of the circular orbit timescale at
$r_{200}$ (which is of the order of the age of the Universe),
$\kappa=t_{\rm relax}/t_{\rm circ}(r_{200})$, the relaxation time may
be written as:
\begin{equation}
\kappa(r)=
\frac{N}{8 \ln N}
\frac{r/V_c}{r_{200}/V_{200}}=
\frac{\sqrt{200}}{8}
\frac{N(r)}{\ln N(r)}
\left(\frac{\overline{\rho}(r)}{\rho_{\rm crit}}\right)^{-1/2},\label{EqnPower}
\end{equation}
where $N=N(r)$ is the enclosed number of particles and ${\overline
\rho}(r)$ is the mean enclosed density within $r$. 

According to P03, deviations of roughly $10\%$ are expected in the
$V_c$ profile where $\kappa \approx 1$, and they adopted this condition
to define the convergence radius, $r_{\rm conv}$. Stricter convergence
demands larger values of $\kappa$, and we shall use a superscript on
$r_{\rm conv}$ to denote the value of $\kappa$ adopted for its
definition. For instance, $r_{\rm conv}^{(1)}=r_{\rm
conv}^{(\kappa=1)}$ corresponds to $\kappa=1$.

Profiles in Fig.~\ref{FigRhoVc} are thus plotted in the range [$r_{\rm
conv}^{(1)}$, $3\, r_{200}$]. As shown in the bottom right panel, this
inner radius indeed corresponds to the point where systematic deviations in
$V_c(r)$ reach $\sim 10\%$. It is also clear from this figure that
convergence in the local density profile is always much easier to
achieve, so concentrating our analysis on the enclosed mass profile,
or on the circular velocity, is a conservative approach.

Although each halo converges over a different radial range, the
departures from convergence are all similar when expressed in terms of
$\kappa$. This is shown in the top panel in Fig.~\ref{FigConvVc},
where differences in $V_c$ from our highest-resolution halo, Aq-A-1,
are shown as a function of $\kappa$ for the other Aq-A
resimulations. Deviations of $\sim 10\%$ are typical at $\kappa=1$;
convergence to better than $\sim 2.5\%$, on the other hand, requires
$\kappa\approx 7$ (indicated by the right dashed vertical line).

\begin{figure}
\begin{center}
\resizebox{8.5cm}{!}{\includegraphics{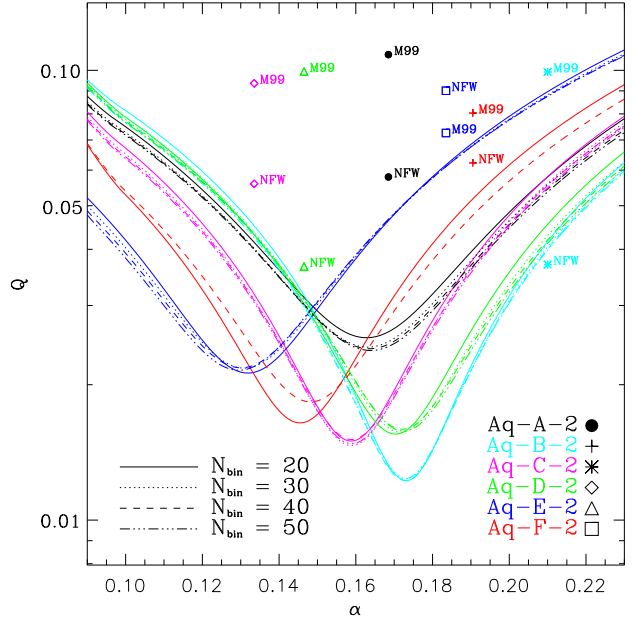}}
\end{center}
\caption{
Minimum-$Q$ values as a function of the Einasto parameter $\alpha$ for
best fits to all level-2 halo profiles in the radial range
$0.01<r/r_{-2}<5$. Colors identify different halos, and line types the
number of bins chosen for the profile. The minimum-$Q$ values obtained
for NFW and M99 best fits are also shown, and are plotted at arbitrary
values of $\alpha$ for clarity. Note that Einasto fits are
consistently better than NFW which are consistently better than M99,
and that a significant improvement in $Q$ is obtained when letting
$\alpha$ vary in the Einasto formula. $Q$ is approximately independent
of the number of bins used in the profile, and is minimized for
different values of $\alpha$ for each individual halo. See text for
further details.
\label{FigLnQAlpha}}
\end{figure}

We may use these results to estimate convergence radii for our highest
resolution run, Aq-A-1: its $V_c$ profile converges to better than
$10\%$ for radii $r>r_{\rm conv}^{(1)}=112 \, h^{-1}$ pc; $2.5\%$
convergence or better is expected for $r> r_{\rm conv}^{(7)}=253 \,
h^{-1}$ pc (see bottom panel of Fig.~\ref{FigConvVc}). Convergence
radii for various values of $\kappa$ are listed in
Table~\ref{tab:FitPar} for each simulated halo.

\begin{figure}
\begin{center}
\resizebox{8.5cm}{!}{\includegraphics{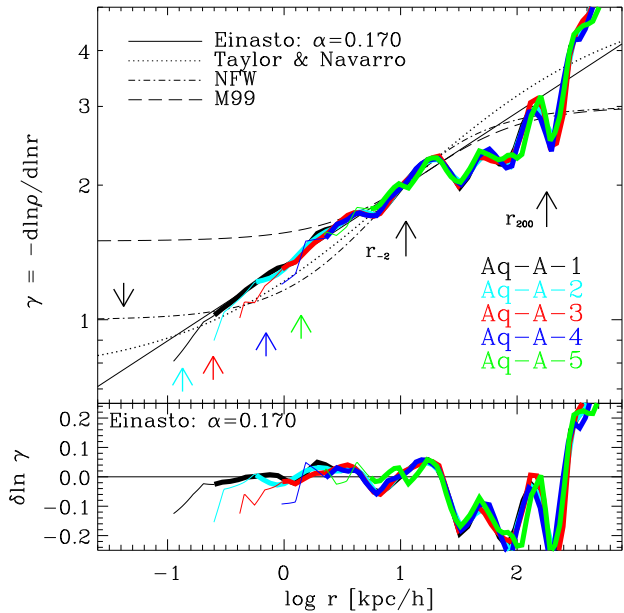}}
\end{center}
\caption{
Logarithmic slope of the density profile as a function of radius for
our Aq-A convergence series.  As in other plots, thick lines show
results for $r>r_{\rm conv}^{(7)}$, thin lines extend the profiles down to
the less strict convergence radius $r_{\rm conv}^{(1)}$. Comparison
shows that excellent numerical
convergence for the slope is achieved down to a radius intermediate
between these two convergence radii. Applied to the highest-resolution
Aq-A-1 simulation, this implies that the slope is shallower than the
asymptotic value of the NFW profile ($r^{-1}$) in the inner
regions. We see no sign of convergence to an asymptotic inner
power-law. Instead, the profiles get shallower toward the centre as
predicted by the Einasto formula (a straight line in this plot). The
``critical solution'' of \citet[]{Taylor2001} (which has a $r^{-0.75}$
asymptotic inner cusp) does better than NFW but not as well as Einasto
in reproducing the inner profile of the halo.
\label{FigDlnrhoDlnr}}
\end{figure}

\begin{figure}
\begin{center}
\resizebox{8.5cm}{!}{\includegraphics{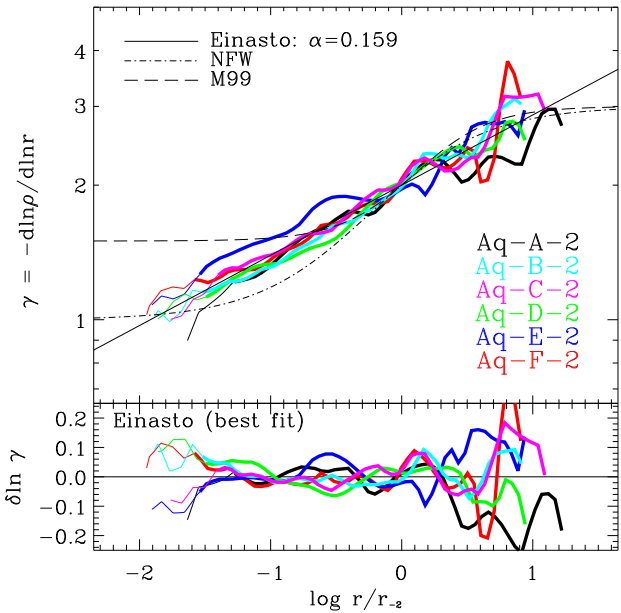}}
\end{center}
\caption{
As Fig.~\ref{FigDlnrhoDlnr}, but for all level-2 resolution
Aquarius halos, after scaling radii to $r_{-2}$. 
\label{FigDlnrhoDlnrProf}}
\end{figure}

\subsection{Fitting formulae}
\label{ssec:fitform}

The fitting formulae we have used to describe the mass profile of our
simulated halos are the following: (i) The NFW profile, given by
\begin{equation}
\rho(r)=
\frac{\rho_s}{(r/r_s)(1+r/r_s)^2},
\label{eq:nfw}
\end{equation}
(ii) the modification to the NFW profile proposed by M99,
\begin{equation}
\rho(r)=
\frac{\rho_M}{(r/r_M)^{1.5}[1+(r/r_M)^{1.5}]},
\label{eq:m99}
\end{equation}
and (iii) the Einasto profile,
\begin{equation}
\ln (\rho(r)/\rho_{-2})=(-2/\alpha)[(r/r_{-2})^{\alpha}-1].
\label{eq:einasto}
\end{equation}

Because each of these formulae defines the characteristic parameters
in a slightly different way, we choose to reparametrise them in
terms of $r_{-2}$ and $\rho_{-2}\equiv \rho(r_{-2})$, which identify
the ``peak'' of the $r^2\rho$ profile shown in the left panel of
Fig.~\ref{FigRhoVc}.  This marks the radius where the logarithmic
slope of the profile, $\gamma(r)=-d\ln \rho/d \ln r$, equals the
isothermal value, $\gamma=2$.

The characteristic radius, $r_{-2}$, is a well-defined scalelength
which is relatively easy to identify in each halo without resorting to
any particular fitting formula. In practice, we determine $r_{-2}$ by
computing the logarithmic slope profile, $\gamma(r)$, and identifying
where a low-order polynomial fit to it intersects the isothermal
value. Each $r^2\, \rho$ profile is then visually inspected in order
to ensure that $r_{-2}$ corresponds to the main peak of the profile,
and that it is not unduly influenced by secondary peaks that arise as
the result of substructure. (See the left panel of
Fig.~\ref{FigRhoVc}.) Table~\ref{tab:FitPar} lists $r_{-2}$ and
$\rho_{-2}$ for all our simulated halos. Note that for the NFW
profile, $r_{-2}=r_s$ and $\rho_{-2}=\rho_s/4$, while for the Moore
profile, $\rho_{-2}=(4/3)\, \rho_M$ and $r_{-2}=2^{-2/3}\, r_M$.

We note that, unlike NFW or M99, when $\alpha$ is allowed to vary
freely the Einasto profile is a 3-parameter fitting formula. This is
not, of course, the only possible extension of NFW-like profiles which
allows for a variable shape with the aid of an extra free
parameter. For example,  \citet{Merritt2006} compared N-body halos with the
3-parameter Einasto formula, as well as  with the anisotropic model of
\citet{Dehnen2005} and with the deprojected \citet{Sersic1968} model
of \citet{Prugniel1997}. Merritt et al.  conclude that, overall,
Einasto's formula performs best. Therefore, we adopt it here for the
rest of our analysis, although we do not exclude the possibility that
other 3-parameter formulae may perform at least as well as
Einasto's. A full exploration of this issue is beyond the scope of
this paper.

\subsection{Fitting procedure}

Best-fit parameters are found by minimizing the deviation between
model and simulation across all bins in a specified radial range.  In
the case of the density profile, the best fit is found by minimizing
the figure-of-merit function, $Q^2$, defined by 
\begin{equation} 
Q^2={1 \over N_{\rm bins}} \sum_{i=1}^{N_{\rm bins}}
(\ln{\rho_i} - \ln{\rho_i^{\rm model}})^2.  \label{eq:q2}
\end{equation} 

This function provides an intuitively simple measure of the level of
disagreement between simulated and model profiles. It is
dimensionless; it weights different radii logarithmically; and, for
given radial range, $Q^2$ is approximately independent of the number
of bins used in the profile.  Thus, minimizing $Q^2$ yields for each
halo well-defined estimates of a model's best-fit parameters. Note
that when $Q$ is small, it is just the {\it rms} fractional deviation
of the data from the model.

It is less clear how to define a goodness-of-fit measure associated
with $Q^2$ and, consequently, how to assign statistically-meaningful
confidence intervals to the best-fit parameter values.  This
difficulty arises because, at the very high resolution of the
simulations analyzed here, discreteness noise in the binned density
estimates is negligible. The figure of merit of a fit therefore
depends not only on how faithfully a model approximates a halo but
also on the presence of individual halo features that no simple
fitting formula can hope to reproduce.  These distinct features are
present on small scales (substructure) and large scales \cite[such as
streams, asphericity, and other relics of each halo's specific
assembly history; see, e.g.,][]{Vogelsberger2009}. As a result,
bin-to-bin residuals are distinctly non-Gaussian and highly
correlated, precluding the use of simple statistical tools such as the
$\chi^2$ distribution in order to assess goodness of fit.

Assessing the acceptability of various $Q$ values would require the
definition of a detailed statistical model in order to measure
reliably the departures of individual halos from a smooth profile
whose average shape (and scatter) could be obtained directly by
averaging various numerical realizations of halos of the same
mass. Unfortunately, such procedure is unlikely to be robust with only
6 halos in our sample.

Therefore, we limit our analysis to comparing the minimum-$Q$ values
obtained with various formulae, and to discussing how $Q$ changes as
the fitting parameters are varied. The actual value of $Q$ is, after
all, a reliable and objective measure of the average per-bin deviation
from a particular model. As we discuss below, this is, in many cases,
enough to prefer unequivocally one fitting formula over another and to
make a compelling case for the need of an extra parameter in the fit.

\subsection{Einasto vs NFW vs M99}

The left panel of Fig.~\ref{FigDensProf} compares the density profiles
of all six level-2 Aquarius halos, after scaling radii to $r_{-2}$ and
densities to $\rho_{-2}$.  The right-hand panel shows the circular
velocity profiles, scaled in an analogous manner to match the peak of
the profile, identified by $r_{\rm max}$ and $V_{\rm max}$. In these
scaled units, the fitting formulae introduced in
Sec.~\ref{ssec:fitform} are curves of fixed shape and normalization,
as shown by the thin solid, dashed, and dot-dashed curves in
Fig.~\ref{FigDensProf}. (The Einasto curve adopts $\alpha=0.159$ in
this figure.)

Comparison with the simulations (thick curves) indicates that there is
a clear mismatch between the shape of the halo profiles and those of
the NFW and M99 fitting formulae. This is not just a result of
enforcing the $r_{-2}$-$\rho_{-2}$ scaling. We illustrate this by
showing, in the two bottom panels of Fig.~\ref{FigDensProf}, residuals
from {\it best fits} obtained by adjusting {\it both} fit parameters
of the NFW and M99 profiles ($r_{-2}$ and $\rho_{-2}$) in order to
minimize $Q^2$. (The radial range chosen for these fits is $r_{\rm
conv}^{(1)} <r<0.5 \, r_{200}$.)  Note the ``S'' shape in the
residuals, which are largest (and increasing) at the innermost radius
of the profile. Because of the shape mismatch, extrapolating either
the NFW or M99 fits further inwards, to regions less well resolved
numerically, is almost guaranteed to incur substantial error.

The large-scale radial trend of the residuals from the best Einasto
fits (middle panels of Fig.~\ref{FigDensProf}), on the other hand, is
rather weak, suggesting that the shape of the simulated halo profiles
are much better accommodated by this formula.  This is {\it not} just
a result of the extra shape parameter in the Einasto formula: even
when keeping $\alpha$ fixed to a single value, residuals are smaller
and have less radial structure than those from either NFW or M99.

We show this in Fig.~\ref{FigLnQAlpha}, where we plot the minimum-$Q$
($Q_{\rm min}$) values of the best Einasto fits for all six level-2
Aquarius halos, as a function of the shape parameter $\alpha$. For
given value of $\alpha$ the remaining two free parameters of the
Einasto formula are allowed to vary in order to minimize
$Q^2$. Different line types correspond to different numbers of bins
used to construct the profile (from $20$ to $50$), chosen to span in
all cases the same radial range, $0.01<r/r_{-2}<5$, a factor of $500$
in radius. Minimum-$Q$ values are computed using a similar procedure
for the NFW and M99 formulae, and are shown, for each halo, with
symbols of corresponding colour.

In terms of $Q_{\rm min}$, Einasto fits are consistently superior to
NFW or M99, whether or not the $\alpha$ parameter is adjusted
freely. For example, for {\it fixed} $\alpha=0.15$, {\it all} Einasto
best fits have minimum-$Q$ values below $\sim 0.03$. For
comparison, best NFW and M99 fits have an average $\langle Q_{\rm min}
\rangle \sim 0.06$ and $0.095$, respectively. These numbers correspond
to $N_{\rm bins}=20$, but they are rather insensitive to $N_{\rm
bins}$, as may be judged from the small difference between the various
lines corresponding to each halo in Fig.~\ref{FigLnQAlpha}.

We emphasize that, although the improvement obtained with Einasto's
formula is significant, NFW fits are still excellent, with a typical
rms deviation of just $\sim 6\%$ over a range of $500$ in radius. The
use of the NFW formula may thus be justified for applications where
this level of accuracy is sufficient over this radial range.

When $\alpha$ is adjusted as a free parameter, $\langle Q_{\rm min}
\rangle \sim 0.018$ for Einasto fits. Furthermore, there is, for each
halo, a well defined value of $\alpha$ that yields an absolute minimum
in $Q$. The $Q$-dependence on $\alpha$ about this minimum is roughly
symmetric and, as expected, nearly independent of the number of bins
used in the profile. The minimum in $Q$ is sharp; a shift of just
$0.015$ in $\alpha$ typically leads to an increase of $\sim 50\%$ in
$Q$ around the minimum. Given that the value of $\alpha$ that
minimizes $Q$ varies from $0.130$ for Aq-E-2 to $0.173$ for Aq-B-2, we
conclude that the improvement obtained when allowing $\alpha$ to vary
is significant. We quote nominal ``error bars'' for $\alpha$ in
Table~\ref{tab:FitPar} that bracket the interval where $Q$ deviates by
less than $50\%$ from the absolute minimum in Fig.~\ref{FigLnQAlpha}.

\subsection{Self-similarity?}

The need for a variable $\alpha$ discussed above illustrates one of
our main findings: namely, that the mass profiles of our Aquarius
halos {\it are not strictly self similar}. The shapes of the profiles
are subtly but significantly different from each other, and no
rescaling can match one exactly to another. Halo Aq-E-2 provides the
most striking example, deviating from halo Aq-D-2, for example, by
almost a factor of $2$ in density at $\sim 0.03 \, r_{-2}$. The same
differences in mass profile shape are also easily appreciated in the
scaled circular velocity profiles, which indicate that the departures
from similarity are genuine and not just caused by inaccuracies in the
scaling or by the ``bumps and wiggles'' caused by unrelaxed tidal
debris and substructure.

We have verified this further by performing the same analysis after
removing bound substructure clumps identified by {\small SUBFIND}: the
same conclusion applies to the ``cleaned'' profiles of the main smooth
halo. With hindsight, this is perhaps not too surprising.  Bound
substructures do not amount to more than $\sim 10\%$ of the halo mass
\citep{AqPaper2}, and therefore cannot alter the results  discussed above.



We have also checked that the differences in $\alpha$ are not caused
by transient departures from equilibrium or numerical resolution: the
same qualitative trends, and indeed very similar $\alpha$ values, are
seen at earlier times and in runs with fewer particles. There also
seems to be little correlation between $\alpha$ and the overall
triaxiality of the system; however, we shall only deal here
with spherically-averaged profiles, and defer a detailed study of
departures from sphericity to a later paper.

Although the departures from similarity appear significant, we must
also emphasize that they are rather subtle, and are only clearly
evident because of the large radial range resolved by our simulations,
about {\it three decades} in radius within the virialized region of a
halo. Simulations with more limited numerical resolution have hinted
at this but had difficulty making such a compelling case for
non-similarity \citep[see,
  e.g.,][]{Navarro2004,Merritt2005,Stoehr2006,Merritt2006}.

\begin{figure}
\begin{center}
\resizebox{8.5cm}{!}{\includegraphics{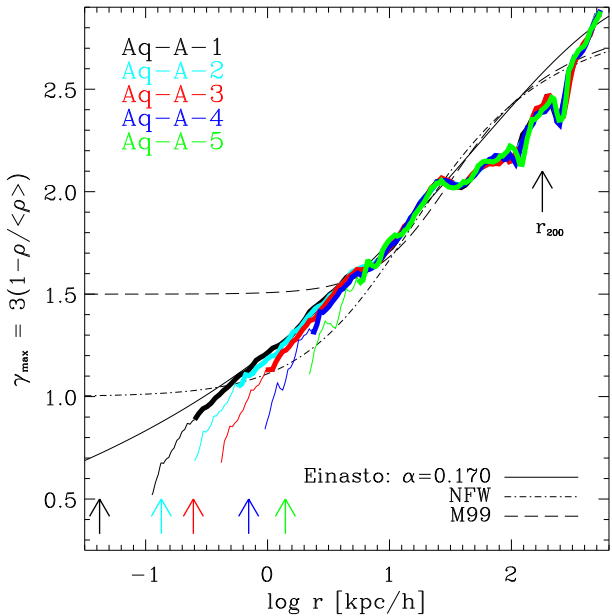}}
\end{center}
\caption{ Maximum value of the asymptotic inner slope of the density
  cusp, as a function of radius for our Aq-A convergence
  series. Excellent numerical convergence is achieved at radii
  comparable to $r_{\rm conv}^{(7)}$ (the inner limit of the thick
  lines; thin lines extend down to $r_{\rm conv}^{(1)}$). This shows
  that there is not enough mass near the centre of Aq-A to sustain a
  cusp steeper than $\rho \propto r^{-0.9 \pm 0.1}$. Arrows are as in
  Fig.~\ref{FigRhoVc}.
\label{FigMaxSlope}}
\end{figure}

\begin{figure}
\begin{center}
\resizebox{8.5cm}{!}{\includegraphics{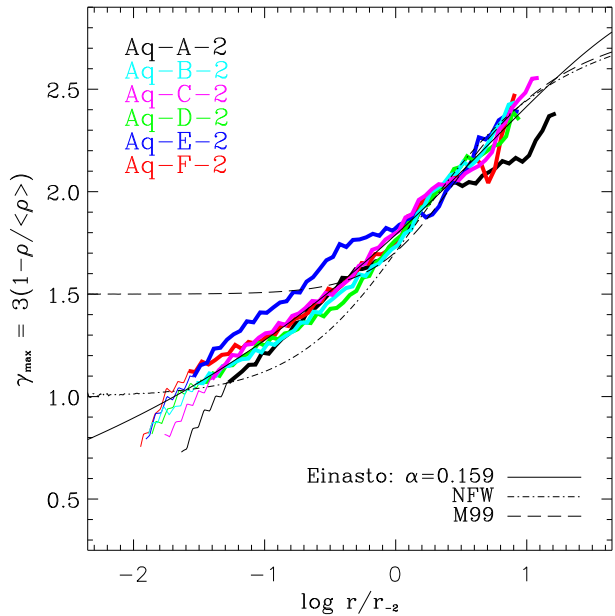}}
\end{center}
\caption{
As Fig.~\ref{FigMaxSlope}, but for our six level-2 Aquarius
halos. Results are similar in all cases and rule out cusps
steeper than $r^{-1}$ for $\Lambda$CDM halos.
\label{FigMaxSlopeProf}}
\end{figure}

\subsection{The Cusp}
\label{ssec:cusp}

It is clear from the residuals in the bottom panels of
Fig.~\ref{FigDensProf} that, near the centre, the M99 profile
approximates the simulated halos more poorly than either NFW or
Einasto. The weak performance of the M99 formula may be traced to its
steep asymptotic inner slope, $\rho \propto r^{-1.5}$. Indeed, all six
Aquarius halos have {\it measured} slopes in the inner regions that
are substantially shallower than $-1.5$.  This is shown in
Figs.~\ref{FigDlnrhoDlnr} and \ref{FigDlnrhoDlnrProf}, where the thick
portion of each curve corresponds to $r>r_{\rm conv}^{(7)}$ and the
innermost point plotted to $r_{\rm conv}^{(1)}$. In all cases, the
logarithmic slopes converge well inside $r_{\rm conv}^{(7)}$, and only
minor deviations may be seen at radii beyond $r_{\rm conv}^{(1)}$.

Interestingly, the slope of the Aq-A-1 profile at $r=r_{\rm
  conv}^{(7)}$ is exactly $-1$, and becomes shallower inward, so it is
clear that at least for this halo we are able to resolve a region
where the dark matter profile has become shallower than $-1$, the
asymptotic value of the NFW profile. Fig.~\ref{FigDlnrhoDlnrProf}
shows the radial dependence of the logarithmic slope for all six
level-2 halos and confirms the general applicability of the Aq-A
results: the {\it measured} slopes of all halos approach $-1$ (and are
certainly shallower than $-1.5$) at the innermost resolved point.

Figs.~\ref{FigDlnrhoDlnr} and ~\ref{FigDlnrhoDlnrProf} also make clear
that there is no sign that the profiles are approaching power-law
behaviour near the centre: they keep getting shallower to the
innermost resolved radius. This behaviour is well captured by the
Einasto model, where the logarithmic slope is simply a power-law of
radius, $d\ln\rho/d\ln r \propto r^{\alpha}$. Our results thus rule
out recent claims of cusps as steep as $r^{-1.2}$ in typical
$\Lambda$CDM halos \citep{Diemand2004,Diemand2005,Diemand2008}. 

This conclusion is unlikely to depend on the details of our profile
construction and/or fitting procedures. Indeed, as we show in the next
subsection, there is actually {\it not enough mass} within the
innermost resolved radius to allow for a cusp as steep as
$r^{-1.2}$. Recent work by \citet{Stadel2008}, also based on very
high-resolution simulations, agrees with our present conclusions, and
argues for asymptotic inner slopes shallower than $-1$, as previously
suggested by \citet{Navarro2004}.

\subsection{The Asymptotic Inner Slope}

The results presented above do not preclude the possibility that a
shallow power-law cusp may be present in the innermost regions which
are still unresolved in our simulations. It is therefore interesting
to estimate the {\it maximum} value that the slope of such a cusp may
take. This is constrained, at any radius, by the total enclosed mass
and the local value of the spherically averaged density: slopes
steeper than $\gamma_{\rm max}$ require more mass than is available
within that radius. This constraint assumes only that the logarithmic
slope is monotonic with radius and that the halo is not hollow. It is
then straightforward to show that the maximum possible inner
asymptotic slope is $\gamma_{\rm max}=3(1-\rho(r)/{\bar \rho}(r))$,
where ${\bar \rho}(r)$ is the mean density enclosed within
$r$. Evaluated at the innermost radius where both local density and
enclosed mass (or, equivalently, circular velocity) have converged,
this quantity provides an important constraint on the density profile
at radii that remain unresolved even in our best simulations.

We show this parameter as a function of radius for our Aq-A
convergence series in Fig.~\ref{FigMaxSlope}. This figure shows that
$\gamma_{\rm max}$ converges to better than 0.1 for $r>r_{\rm
  conv}^{(7)}$ (the innermost point of the thick portion of the
profiles). Our data for Aq-A thus indicates that there is not enough
mass in the unresolved region to support a cusp steeper than
$r^{-0.9\pm0.1}$. Fig.~\ref{FigMaxSlopeProf} shows that the results
for Aq-A are not exceptional: all our level-2 Aquarius
halos suggest maximum possible asymptotic slopes of about $-1$.

\begin{figure*}
\begin{center}
\resizebox{17.5cm}{!}{\includegraphics{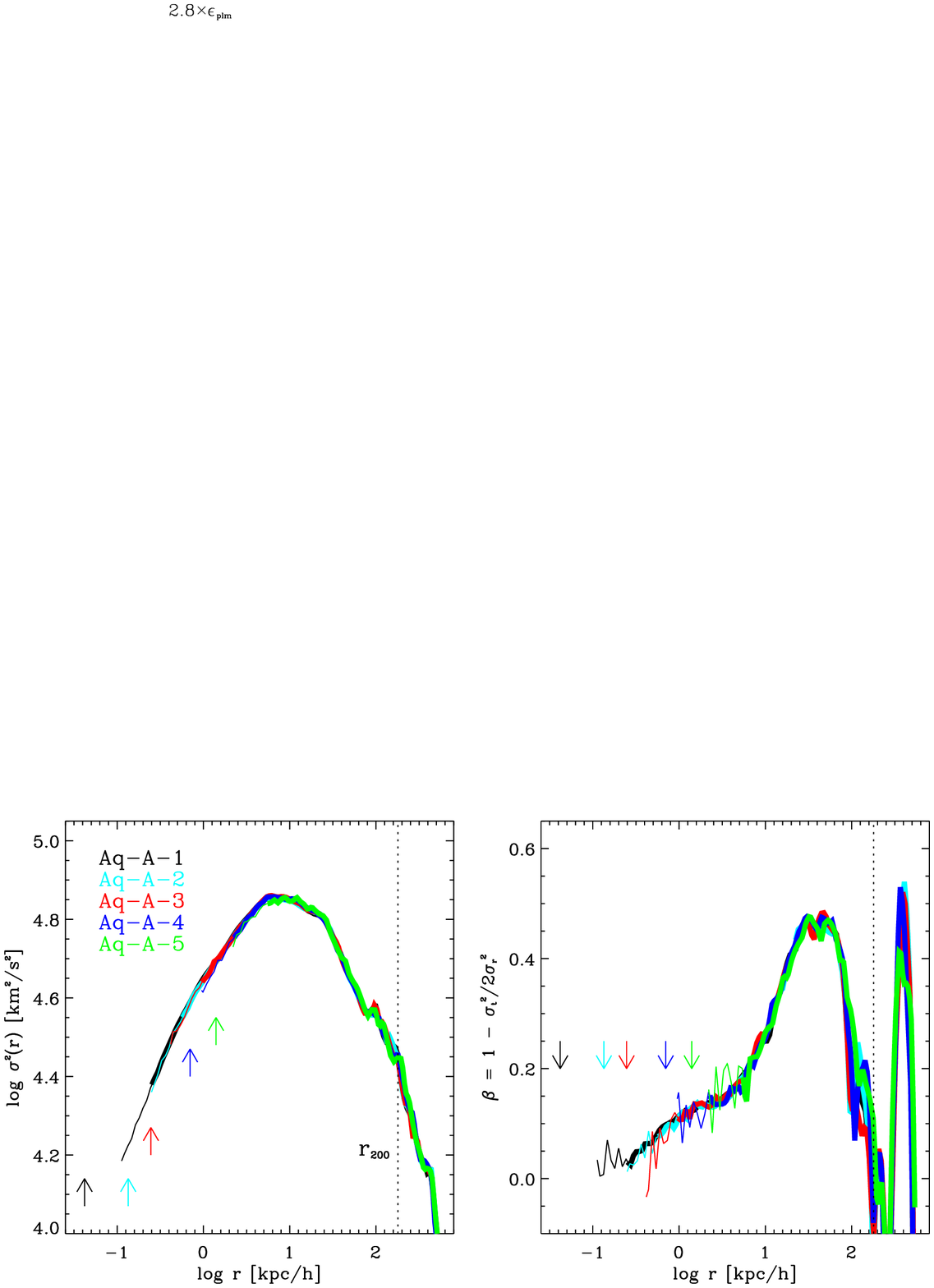}}
\end{center}
\caption{ {\it Left panel:} Velocity dispersion profiles for our Aq-A
  convergence series. Arrows, line-types and colours are as in
  Fig.~\ref{FigRhoVc}. Note the excellent numerical convergence. The
  shape of the velocity dispersion profile is remarkably similar to
  that of the $r^2 \rho$ profile shown in Fig.~\ref{FigRhoVc},
  highlighting the intimate connection between the density and
  velocity dispersion profiles which is responsible for the power-law
  behaviour of the pseudo-phase-space density profile discussed in
  Sec.~\ref{ssec:pdprof}. {\it Right panel:} Anisotropy profiles for
  the Aq-A convergence series. Note the non-monotonic variation with
  radius: the halo is nearly isotropic near the centre, is dominated
  by radial motions at intermediate radii, but becomes markedly less
  anisotropic near the virial radius.
\label{FigVelDispAnis}}
\end{figure*}
\begin{figure*}
\begin{center}
\resizebox{15cm}{!}{\includegraphics{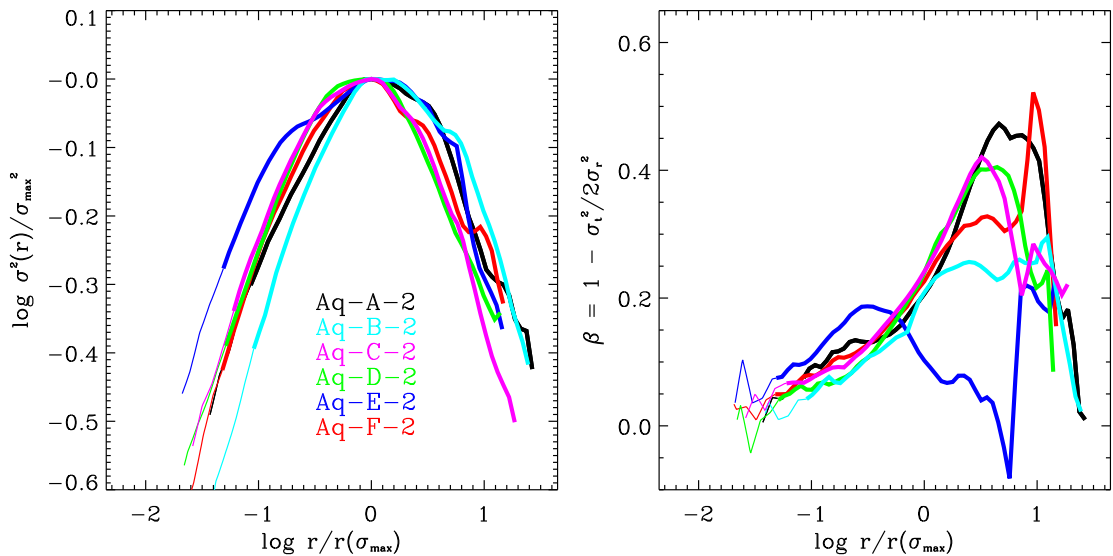}}
\end{center}
\caption{
As Fig.~\ref{FigVelDispAnis}, but for all six level-2 resolution
Aquarius halos, scaled to match at the peak of the profile, identified
by $\sigma_{\rm max}$ and $r(\sigma_{\rm max})$. This scaling
highlights small but significant departures from similarity in the
velocity dispersion structure of $\Lambda$CDM halos.  Note the correspondence in
shape between the velocity dispersion and $r^2\rho$ profiles shown in
Fig.~\ref{FigRhoVc}, which reflects the ``universal'' pseudo-phase-space density
profile of the halos (Fig.~\ref{FigPhaseDensProf}). Note also that
the non-monotonic behaviour of the anisotropy highlighted in
Fig.~\ref{FigVelDispAnis} is common to all six halos.
\label{FigVelDispAnisProf}}
\end{figure*}

\section{Dynamical Profiles}
\label{sec:dynprof}

\subsection{Velocity Dispersion Structure}
\label{ssec:velprof}

Fig.~\ref{FigVelDispAnis} shows velocity dispersion and anisotropy
profiles for our Aq-A series and demonstrates that the excellent
numerical convergence of our simulations extends to their velocity
dispersion structure. The velocity dispersion (squared) is computed
simply as twice the specific kinetic energy in each spherical shell
and the anisotropy as $\beta=1-\sigma_t^2/(2\sigma_r^2)$, where
$\sigma_t^2$ and $\sigma_r^2$ are the (squared) velocity dispersion in
tangential and radial motions, respectively.  Besides numerical
convergence, the panels in this figure illustrate two important
points. The first concerns the shape of the velocity dispersion
profiles (left panel in Fig.~\ref{FigVelDispAnis}), which is
remarkably similar to that of the $r^2\rho$ profiles shown in
Fig.~\ref{FigRhoVc}. This coincidence suggests an intimate connection
between density and velocity dispersion, which we explore in more
detail in Sec.~\ref{ssec:pdprof}. The second point concerns the
anisotropy profile, which is clearly non-monotonic. It is nearly
isotropic at the centre, becomes radially anisotropic at intermediate
radii, but the dominance of radial motions decreases again near the
virial radius. As shown in Fig.~\ref{FigVelDispAnisProf}, these
properties appear to be rather general, since all six Aquarius halos
have non-monotonic anisotropy profiles and similar velocity dispersion
profile shapes.

\subsection{Self-similarity?}

Fig.~\ref{FigVelDispAnisProf} also demonstrates a clear lack of
self-similarity in the structure of the simulated halos. We have
chosen to emphasize this by rescaling all profiles so as to match the
peak of the $\sigma(r)$ curve, which occurs at $r(\sigma_{\rm
max})$. This scaling demonstrates that, as with the density profiles, the
{\it shape} of the $\sigma(r)$ profiles differs subtly but significantly
amongst halos. We have checked that these differences in shape are
not due to bound subhalos; removing all the subhalos identified
by our SUBFIND algorithm and recalculating the dispersion and
anisotropy  profiles results in only rather minor changes

The most striking case is again that of halo Aq-E-2 (blue curve),
whose $\sigma(r)$ profile is much broader than the others. Recall that
this halo also stands out in Fig.~\ref{FigDensProf} as having an
unusually broad $r^2\rho$ profile. Halo Aq-E-2 also has an unusual
velocity anisotropy profile, with less predominance of radial motions
than the rest of the series. The departures from similarity in mass
and velocity structure therefore seem closely linked, suggesting that
these halos may share a common property that combines density and
velocity dispersion. We explore this in Sec.~\ref{ssec:pdprof} below.

\subsection{Anisotropy-slope relation}

\begin{figure}
\begin{center}
\resizebox{8.5cm}{!}{\includegraphics{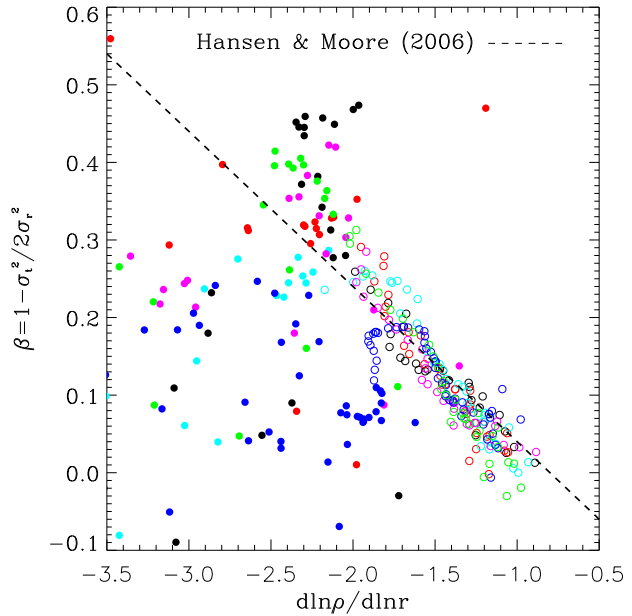}}
\end{center}
\caption{ Local values of the logarithmic slope of the density profile
  plotted versus velocity anisotropy. The relation proposed by
  \citet[]{Hansen2006} is shown as a dashed line. Because the density
  profile steepens gradually from the centre outwards whereas the
  velocity anisotropy is non-monotonic, no simple relation between
  these two quantities is valid throughout the halos. The Hansen \&
  Moore formula approximates our results quite well in the inner
  regions, but large deviations may be seen outside $r_{-2}$,
  particularly at the largest radii where our halos are approximately
  isotropic but their density profiles are steepest. Open circles
  correspond to $r_{\rm conv}^{(1)}<r<r_{-2}$, filled circles to
  $r_{-2}<r<r_{200}$. Colors are as in
  Fig.~\ref{FigDensProf}. \label{FigBetaSlope}}
\end{figure}

We may use the results of the previous subsection to assess recent
claims by \citet[]{Hansen2006} of a general connection between the
local values of logarithmic slope, $\gamma$, and the velocity
anisotropy, $\beta$. We show this in Fig.~\ref{FigBetaSlope}, where we
plot $\beta$ vs $\gamma$ for all level-2 Aquarius halos. Open circles
correspond to the inner regions of the halo ($r_{\rm
conv}^{(1)}<r<r_{-2}$) whereas filled circles correspond to the outer
regions ($r_{-2}<r<r_{200}$). As in other figures, different colours
correspond to the different Aquarius halos. The relation proposed by
Hansen \& Moore is shown by a dashed line and accounts reasonably well
(albeit not perfectly) for our data in the inner regions where both
the anisotropy and the logarithmic slope are monotonic functions of
$r$.

However, there are large departures from this relation in the outer
regions, where the density profile steepens further but the velocity
ellipsoid tends to become less anisotropic. The failure of the Hansen
\& Moore relation in the outer regions is not unexpected since
$\gamma$, unlike $\beta$, is monotonic with radius. We conclude that,
if a simple relation links anisotropy and slope, it can only hold in
the inner regions of halos.
\subsection{The Phase-Space Density Profile}
\label{ssec:pdprof}

The similarity in shape between the $\sigma^2$ and $r^2\rho$ profiles
highlighted above suggests that there may be a simple scaling between
densities and velocity dispersions in halos. This is best appreciated
by considering the quantity $\rho/\sigma^3$, which, for dimensional
reasons, we shall call the pseudo-phase-space density, although it is
important to realise that it is {\it not} the true coarse-grained
phase-space density at the resolution of our simulations, or even the
average of this quantity in spherical shells. For consistency with the
rest of our analysis, we calculate $\rho/\sigma^3$ directly from the
estimates of $\rho$ and $\sigma$ computed in concentric spherical
shells.

\begin{figure*}
\begin{center}
\resizebox{17.5cm}{!}{\includegraphics{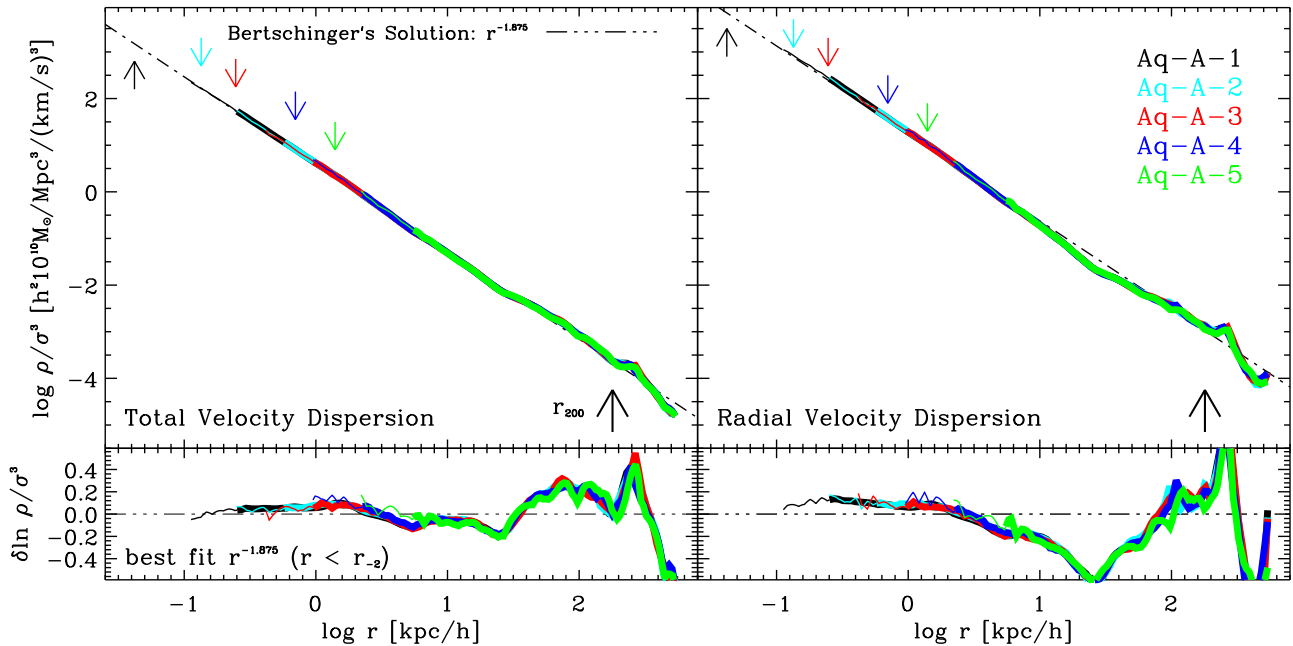}}
\end{center}
\caption{ Pseudo-phase-space density profiles for our Aq-A convergence
  series, estimated as $\rho/\sigma^3$, computed in concentric
  spherical shells. Arrows, line-types, and colours are as in
  Fig.~\ref{FigRhoVc}. Note the remarkable power-law behaviour of this
  quantity, a result already noted by \citet[]{Taylor2001}. The
  dot-dashed line is not a fit to the data, but rather the prediction
  of the similarity solution of \citet[]{Bertschinger1985} for infall
  onto a point mass in an otherwise unperturbed Einstein-de Sitter
  universe, $\rho/\sigma^3 \propto r^{-1.875}$. This has been scaled
  to match Aq-A at $r<r_{-2}$.  Residuals from the Bertschinger
  solution are shown in the bottom panels. Note that this power-law
  behaviour is most evident when the full 3D velocity dispersion is
  used (left panels). When only the radial velocity dispersion is used
  (right panels) deviations from the Bertschinger solution are
  considerably larger.
\label{FigPhaseDens}}
\end{figure*}

Fig.~\ref{FigPhaseDens} shows the $\rho/\sigma^3$ profile for our Aq-A
convergence series. As noted by \citet[]{Taylor2001}, the 
profile of this quantity is remarkably well approximated by a
power-law. More remarkable still is the fact that the power law is
indistinguishable from that predicted by the similarity solution of
\citet[]{Bertschinger1985} for infall onto a point mass in an
otherwise unperturbed Einstein-de Sitter universe, $\rho/\sigma^3\propto
r^{-1.875}$ (dot-dashed line in Fig.~\ref{FigPhaseDens}). This
solution is spherically symmetric, involves purely radial motions, and
is violently dynamically unstable, so its relevance to $\Lambda$CDM
halos is far from clear. The residuals in the bottom panel of
Fig.~\ref{FigPhaseDens} are deviations from a Bertschinger law matched
within the characteristic radius $r_{-2}$, where substructure
bumps and wiggles are minimal.

Note that, although there is only {\it one free parameter} in this fit
(the vertical scaling), the residuals do not exceed $\sim 20\%$ {\it
  anywhere} within the virial radius, even though substructures add
significant noise to the dynamical measurements in the outskirts of
the halo. Interestingly, the residuals increase when $\sigma_r$, the
velocity dispersion in radial motions, is used in place of the full 3D
rms velocity, $\sigma$,to estimate the ``phase-space density''. Thus,
the $r^{-1.875}$ behaviour seems to concern the full kinetic energy
content of each shell rather than just radial or tangential motions.

\begin{figure*}
\begin{center}
\resizebox{17.5cm}{!}{\includegraphics{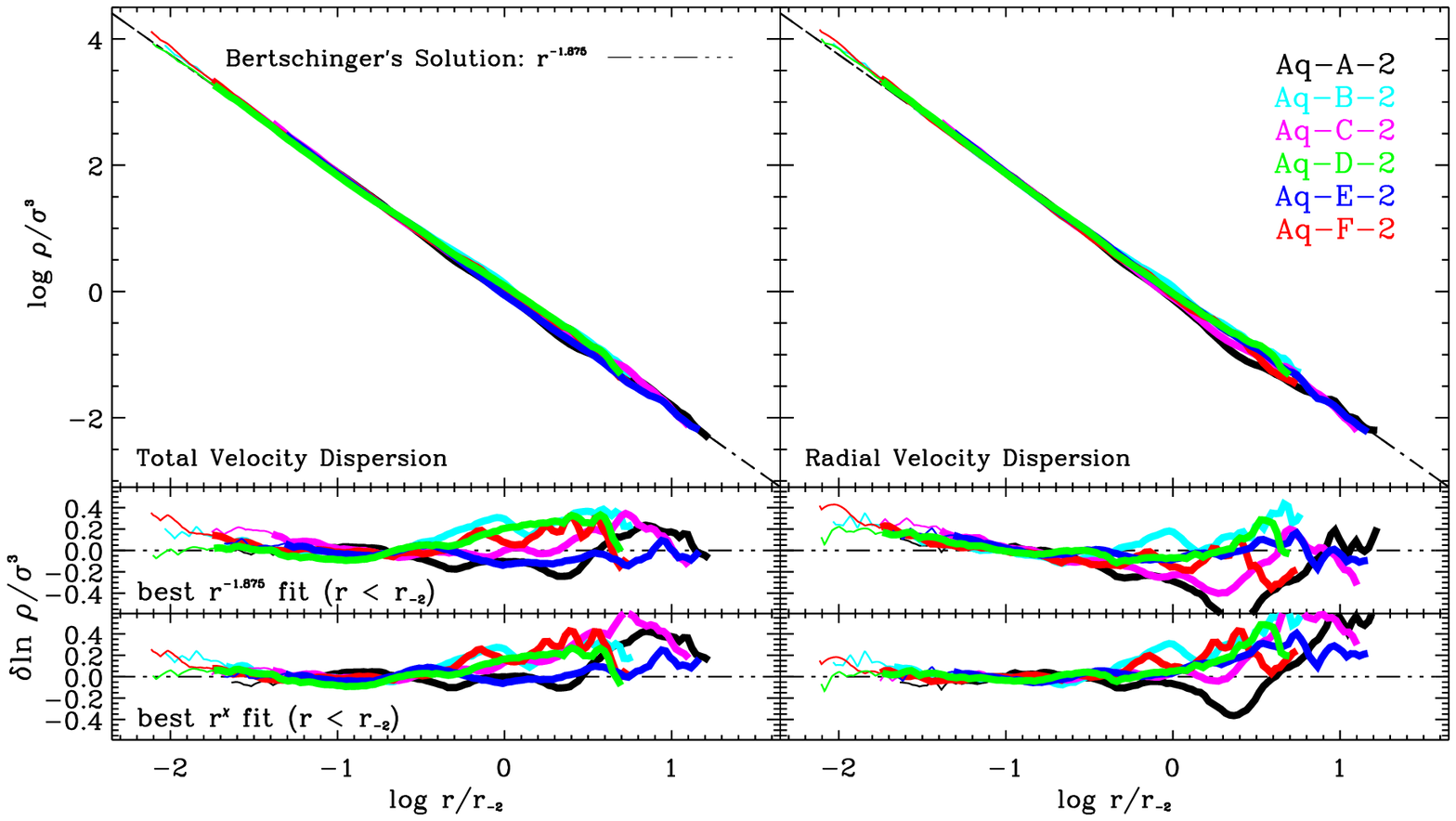}}
\end{center}
\caption{Pseudo-phase-space density profiles of all six level-2 Aquarius
  halos. Radii have been scaled to $r_{-2}$, and the
  pseudo-phase-space densities to maximise agreement within
  $r_{-2}$. Note that for all six halos these profiles are very well
  approximated by power laws with an exponent very close to that of
  the Bertschinger solution. All halos, including those that were
  outliers in the density, velocity dispersion, and anisotropy
  profiles, are almost indistinguishable in this plot.  Deviations
  from the Bertschinger law are typically more pronounced when radial
  velocity dispersion is used instead of the full 3D velocity
  dispersion. Residuals from the best-fit power-laws, $\rho/\sigma^3
  \propto r^{\chi}$, are shown in the bottom panels. The values of
  $\chi$ are listed for each halo in Table~\ref{tab:FitPar}.
\label{FigPhaseDensProf}}
\end{figure*}

Fig.~\ref{FigPhaseDensProf} shows that similar conclusions apply to
the rest of the Aquarius halos. Residuals from the Bertschinger law
are small for all halos, and are typically larger when the radial
velocity dispersion is used. Note that there is some ``curvature'' in
the residual profiles, suggesting that a power-law is a good, but
perhaps not perfect, description of the radial dependence of
$\rho/\sigma^3$. We are currently investigating the origin of this
curvature and plan to report on it in a future paper (Ludlow et al.,
in preparation).

A power-law radial dependence is approximately preserved when
$\sigma_r$ is used, but the best fitting value of the exponent differs
systematically from $-1.875$. This may be seen in the bottom panels of
Fig.~\ref{FigPhaseDensProf}, which show the residuals from the best
fitting $\rho/\sigma^3 \propto r^{\chi}$ law. The values of the
best-fit exponent for both $\rho/\sigma^3$ and $\rho/\sigma_r^3$
($\chi$ and $\chi_r$, respectively) are listed in
Table~\ref{tab:FitPar}.

Perhaps the most important result from Fig.~\ref{FigPhaseDensProf} is
that there seems to be very little scatter between halos when
considering their $\rho/\sigma^3$ profiles. Take, for example, the
case of halo Aq-E-2, which was a clear outlier in the density,
velocity dispersion, and anisotropy profiles. When considering
$\rho/\sigma^3$ this halo is unremarkable, and follows the Bertschinger
law as closely as the others.

This shows that there is a sense in which $\Lambda$CDM halos are
nearly universal, but that universality does not extend to their
density or velocity dispersion profiles separately, but rather only to
their pseudo-phase-space density profile. This may appear a bold
statement, and it certainly needs to be corroborated by future work,
but it offers an intriguing perspective into the origin of the
near-universal density profiles of halos, the meaning of the Einasto
shape parameter, $\alpha$, and the provenance of their velocity
dispersion structure. These issues deserve further investigation.

We end by noting that, although it is still not clear what leads to
the power-law stratification of $\rho/\sigma^3$, these results may be
used to place constraints on the structure of the central cusp, under
the plausible (but admittedly unproven) assumption that the power-law
behaviour of the phase-space density continues all the way to the
centre. For example, \citet[]{Taylor2001} used this assumption to show
that, for isotropic systems, a power-law pseudo-phase-space density
implies an inner density cusp with $\rho \propto r^{-0.75}$. This is
certainly consistent with the results shown in Fig.~\ref{FigMaxSlope},
which only exclude cusps steeper than $r^{-0.9 \pm 0.1}$. However, as
we show in Fig~\ref{FigDlnrhoDlnr}, the detailed profile which they
derive for an isotropic halo with Bertschinger's power-law
$\rho/\sigma^3$ profile is a significantly worse fit to our numerical
data than the Einasto profile.

The power-law behaviour of the pseudo phase-space density has been
confirmed by a number of authors, and seems to be present even at
early redshift \citep{Vass2008}. Interestingly, the average power-law
exponent to the $\rho/\sigma_r^3$ profile is $\langle \chi_r
\rangle\approx 1.97$, close to the ``critical'' $1.94$ required by
\citet{Dehnen2005} to have a dynamical model that is well behaved at
all radii. Simulations of even larger dynamic range seem required in
order to explore the true asymptotic inner behaviour of the dynamical
profile of a halo, if indeed there is any such asymptote.


\section{Summary}
\label{sec:conc}

We have analysed density, velocity dispersion, anisotropy and
pseudo-phase-space density profiles at redshift zero for simulated
halos from the {\it Aquarius Project}. This is a set of six
galaxy-sized halos whose formation and evolution have been simulated
at a variety of resolutions in their proper $\Lambda$CDM context. The
set includes the largest simulation of this kind reported so far; a
$\sim 4.4$ billion particle simulation in which the final halo has 1.1
billion particles within its virial radius, $r_{200}$. The set also
includes simulations of all six halos with 100 -- 200 million
particles within the virial radius, as well as a comprehensive
numerical convergence study for the largest system. Our main
conclusions are as follows.

\begin{itemize}

\item Density profiles deviate slightly but significantly from
  the NFW model, and are approximated well by a fitting formula
  where the logarithmic slope is a power-law of radius: the Einasto
  profile (eq.~\ref{eq:einasto}). The steeply-cusped profile of Moore et al.
  (1999) is a poor fit to the structure of our six halos.

\item We find convincing evidence that the shape parameter of the
  Einasto formula varies from halo to halo at given mass (see
  Table~\ref{tab:FitPar}). This complements the earlier conclusion of
  \citet{Merritt2006}, \citet{Gao2008} and \citet{Hayashi2008} that
  its mean value varies systematically with halo mass. Together these
  results imply that the density profiles of $\Lambda$CDM halos are
  not strictly self-similar: different halos cannot be rescaled to
  look alike. This lack of similarity extends to the kinematic
  structure, as measured by the velocity dispersion and anisotropy
  profiles.

\item Intriguingly, departures from similarity are minimized when
  analyzing a pseudo-phase-space density profile defined as
  $\rho/\sigma^3$. This suggests a limited sense in which $\Lambda$CDM
  halos are indeed nearly ``universal''. The pseudo-phase-space
  density profiles are very well approximated by $\rho/\sigma^3
  \propto r^{-1.875}$, the power law predicted by Bertschinger's
  similarity solution for infall onto a point mass in an otherwise
  unperturbed Einstein-de Sitter universe.  This simple law has only
  {\it one} scaling parameter and no shape parameters, yet it
  approximates, for over six decades, the $\rho/\sigma^3$ profiles to
  better than $20$-$30\%$, all the way from the innermost resolved
  point to the virial radius. The power-law description is, however,
  not perfect, and further work designed to understand better its
  origin and limitations seems warranted.

\item Density profiles become monotonically shallower inwards,
  down to the innermost resolved point, with no indication that they
  approach power-law behaviour. The innermost slope we measure is
  slightly shallower than ${-1}$, a result supported by estimates of
  the maximum possible asymptotic inner slope. 

\item These results convincingly rule out recent claims that typical
  $\Lambda$CDM halos may have asymptotic central cusps as steep as
  $r^{-1.2}$ \citep{Diemand2004,Diemand2005,Diemand2008}. Shallower
  cusps, such as the asymptotic $r^{-0.75}$ behaviour predicted by the
  model of \citet[]{Taylor2001}, cannot yet be excluded. These results
  should discourage further work assuming CDM cusps steeper than
  $r^{-1}$ except possibly around central black holes.

\item Velocity anisotropy does not depend monotonically on
  radius beyond $r_{-2}$. Halos are roughly isotropic near the centre,
  are dominated by radial motions at intermediate radii, but become
  more isotropic again as the virial radius is approached. This
  behaviour does not appear to be driven by the presence of
  substructure. Given that the slope of the density profile does
  increase monotonically with radius, this implies that no simple
  relation between anisotropy and slope can hold throughout a
  halo. The relation recently proposed by \citet[]{Hansen2006} works
  reasonably well in the inner regions ($r<r_{-2}$), but fails at
  larger radii.

\end{itemize}

The main aim of the {\it Aquarius Project} is to provide reliable
theoretical predictions for the structure and formation history of dark
matter halos like that surrounding the Milky Way down to radii of
order $100$ pc. This permits direct comparisons with a number of
observations with minimal extrapolation, and it helps to design new
observational strategies aimed at testing the cold dark matter
paradigm on these very non-linear scales. 

We recognize, however, that many of these tests and predictions will
apply to regions where baryons play an important dynamical role. Our
numerical work provides robust results for the limiting but
unrealistic case of {\it pure dark matter} halos, and these will
undoubtedly be modified in non-trivial ways by the presence of
baryons. Providing a full account of the coupled structure of the cold
dark matter {\it and} baryonic components in galaxies like our own is
clearly the next major computational challenge, and it is likely to
exercise us for some time to come.

\section*{Acknowledgments}
The simulations for the Aquarius Project were carried out at the
Leibniz Computing Center, Garching, Germany, at the Computing Centre
of the Max-Planck-Society in Garching, at the Institute for
Computational Cosmology in Durham, and on the `STELLA' supercomputer
of the LOFAR experiment at the University of Groningen. This work was
supported in part by an STFC rolling grant to the ICC. CSF
acknowledges a Royal Society Wolfson Research Merit award. AH
acknowledges financial support from NOVA and NWO.

\begin{table*}
\center
\begin{tabular}{l l l l l l l l l l l}\hline \hline
Halo       &$m_{\rm p}$     &$\epsilon_{\small G}$&$r_{200}$&   $M_{200}$   &$N_{200}$&$V_{\rm max}$&$r_{\rm max}$&$\sigma_{\rm host}$&$\sigma_{\rm max}$ \\
           &  [$M_{\odot}/h$]  &  [pc$/h$]&[kpc$/h$]&[$M_{\odot}/h$]& [$10^6$]&  [km/s]        &  [kpc$/h$]     &  [km/s]              &  [km/s]      \\ \hline

Aq-A-1     &  1.250$\times 10^3$  &  14  & 179.41   & 1.343$\times 10^{12}$   & 1074.06   & 208.75   &  20.69  & 117.47   & 261.70    \\
Aq-A-2     &  1.000$\times 10^4$  &  48  & 179.49   & 1.345$\times 10^{12}$   & 134.47    & 208.49   &  20.54  & 117.13   & 261.88    \\
Aq-A-3     &  3.585$\times 10^4$  &  87  & 179.31   & 1.341$\times 10^{12}$   & 37.39     & 209.22   &  20.35  & 117.31   & 262.80    \\
Aq-A-4     &  2.868$\times 10^5$  &  250 & 179.36   & 1.342$\times 10^{12}$   & 4.68      & 209.24   &  20.58  & 117.23   & 262.29    \\
Aq-A-5     &  2.294$\times 10^6$  &  500 & 180.05   & 1.357$\times 10^{12}$   & 0.59      & 209.17   &  20.84  & 116.61   & 260.59    \\ \hline

Aq-B-2     &  4.706$\times 10^3$  &  48  & 137.02   & 5.982$\times 10^{11}$   & 127.09    & 157.68   &  29.31  & 89.59    & 190.74    \\
Aq-C-2     &  1.021$\times 10^4$  &  48  & 177.26   & 1.295$\times 10^{12}$   & 126.77    & 222.40   &  23.70  & 124.08   & 270.50    \\
Aq-D-2     &  1.020$\times 10^4$  &  48  & 177.28   & 1.295$\times 10^{12}$   & 126.98    & 203.20   &  39.48  & 113.15   & 254.28    \\
Aq-E-2     &  7.002$\times 10^3$  &  48  & 154.96   & 8.652$\times 10^{11}$   & 123.56    & 179.00   &  40.52  & 101.73   & 215.14    \\
Aq-F-2     &  4.946$\times 10^3$  &  48  & 152.72   & 8.282$\times 10^{11}$   & 167.45    & 169.08   &  31.15  & 96.78    & 204.53    \\
 
\hline
\end{tabular}
\caption{Basic parameters of the Aquarius simulations. We have simulated 6
  different halos, each at several different numerical
  resolutions. The leftmost column gives the simulation name, encoding
  the halo (A to F), and the resolution level (1 to 5; 1 is our
  highest resolution, 5 the lowest).  $m_{\rm p}$ is the particle mass
  in the high-resolution region, $\epsilon_G$ is the
  Plummer-equivalent gravitational softening length, $r_{200}$ is the
  virial radius, defined as the radius enclosing a mean overdensity
  200 times the critical value for closure, $M_{\rm 200}$ is the mass
  within the virial radius, $N_{200}$ is the total number of particles
  within $r_{200}$. Other characteristic properties of the halos
  listed are the position ($r_{\rm max}$) of the peak ($V_{\rm max}$)
  of the circular velocity profile, as well as the 1D velocity
  dispersion of the main halo ($\sigma_{\rm host}$), and the
  peak ($\sigma_{\rm max}$) of the velocity dispersion profile.
  \label{tab:AqSims}}
\end{table*}

\begin{table*}
\center
\begin{tabular}{l l l l l c l l l}\hline \hline
Halo       & $r_{\rm{conv}}^{(1)}$ & $r_{\rm{conv}}^{(7)}$ &$\rho_{-2}$                  & $r_{-2}$ & $\alpha$ & $\chi$ & $\chi_{r}$ & $\gamma_{\rm {max}}$ \\
           & [kpc$/h$]             & [kpc$/h$]             &$[10^{10}h^2M_{\odot}/$Mpc$^3]$&[kpc$/h$] &          &        &            &                      \\ \hline
Aq-A-1     & 0.113 & 0.253 & 7.462$\times 10^5$ & 11.05 & 0.170 $\pm$ 0.0259 & -1.898 & -1.948 & 0.894  \\
Aq-A-2     & 0.250 & 0.575 & 7.322$\times 10^5$ & 11.15 & 0.163 $\pm$ 0.0249 & -1.917 & -1.976 & 1.051 \\
Aq-A-3     & 0.417 & 0.966 & 7.456$\times 10^5$ & 11.09 & 0.174 $\pm$ 0.0266 & -1.926 & -1.995 & 1.128 \\
Aq-A-4     & 0.952 & 2.277 & 6.501$\times 10^5$ & 11.90 & 0.160 $\pm$ 0.0248 & -1.991 & -2.061 & 1.321 \\
Aq-A-5     & 2.206 & 5.530 & 7.534$\times 10^5$ & 11.02 & 0.165 $\pm$ 0.0268 & -2.015 & -2.111 & 1.493 \\ \hline

Aq-B-2 & 0.219 & 0.507 & 1.830$\times 10^5$ & 16.79 & 0.173 $\pm$
0.0123 & -1.868 & -1.938 & 1.039 \\ Aq-C-2 & 0.248 & 0.573 &
4.973$\times 10^5$ & 14.37 & 0.159 $\pm$ 0.0125 & -1.948 & -2.010 &
1.077 \\ Aq-D-2 & 0.281 & 0.652 & 2.075$\times 10^5$ & 20.30 & 0.170
$\pm$ 0.0124 & -1.862 & -1.942 & 1.070 \\ Aq-E-2 & 0.223 & 0.516 &
2.058$\times 10^5$ & 17.88 & 0.130 $\pm$ 0.0200 & -1.912 & -1.947 &
1.084 \\ Aq-F-2 & 0.209 & 0.486 & 1.673$\times 10^5$ & 18.84 & 0.145
$\pm$ 0.0167 & -1.911 & -1.980 & 1.298 \\ \hline \end{tabular}
\caption{ Fit parameters of Aquarius halos. The first column labels
each halo, as in Table~\ref{tab:AqSims}, the second and third list the
convergence radii obtained for $\kappa=1$ and $\kappa=7$. These radii,
$r_{\rm conv}^{(1)}$ and $r_{\rm conv}^{(7)}$, respectively,
correspond to where departures from convergence in the circular
velocity are expected to be of order $10\%$ and $2.5\%$. The
characteristic scale radius $r_{-2}$ corresponds to where the
logarithmic slope equals the isothermal value;
$\rho_{-2}=\rho(r_{-2})$, and $\alpha$ is the best-fit Einasto
parameter. The uncertainty in $\alpha$ indicates the range where
$\Delta Q/Q$ deviates by less than $50\%$ from the absolute minimum
shown in Fig.~\ref{FigLnQAlpha}. Strictly, these are non-symmetric,
so we conservatively quote the largest deviation, positive or
negative. $\chi$ refers to the exponent of the best fitting power-law
describing the $\rho/\sigma^3$ profile. $\chi_r$ is analogous to
$\chi$, but for $\rho/\sigma_r^3$, where $\sigma_r$ is the rms
velocity in radial motions. $\chi$ and $\chi_r$ are computed by
minimizing residuals in the region $r_{\rm
conv}^{(1)}<r<r_{-2}$. Finally, $\gamma_{\rm max}$ lists the value of
the maximum asymptotic slope of the density profile cusp, measured at
$r=r_{\rm conv}^{(7)}$. \label{tab:FitPar}} \end{table*}

\bibliographystyle{mnras}
\bibliography{paper}

\end{document}